\documentclass[12pt]{iopart}
\usepackage{iopams}
\usepackage{graphicx}
\usepackage{hyperref}

\begin{document}
\title[On maximal analytical extension of the Vaidya metric]
{On maximal analytical extension of the Vaidya metric with linear mass function}
\author{V A Berezin\,$^{a}$, V I Dokuchaev\,$^{a,b}$ and Yu N Eroshenko\,$^{a}$}
\address{$^a\,$\scriptsize Institute for Nuclear Research of the Russian Academy of Sciences,
60th October Anniversary Prospect 7a, 117312 Moscow, Russia}
\address{$^b\,$\scriptsize National Research Nuclear University MEPhI (Moscow
Engineering Physics Institute), 115409 Moscow, Russia}
\eads{\mailto{berezin@inr.ac.ru}, 
	\mailto{dokuchaev@inr.ac.ru}, 
	\mailto{eroshenko@inr.ac.ru}}

\begin{abstract} 
The classical Vaidya metric is transformed to the special diagonal coordinates in the case of the linear mass function allowing rather easy treatment. We find the exact analytical expressions for metric functions in this diagonal coordinates. Using these coordinates, we elaborate the maximum analytic extension of the Vaidya metric with a linear growth of the black hole mass and construct the corresponding Carter-Penrose diagrams for different specific cases. The derived global geometry seemingly is valid also for a more general behavior of the black hole mass in the Vaidya metric.
\end{abstract}
%\submitto{\CQG}
%\maketitle
%\tableofcontents

\section{Introduction}

The nonstationary solution of the Einstein equations, derived by Prahalad Chunnilal Vaidya \cite{Vaidya43,Vaidya51,Vaidya53}, has a very simple form
\begin{equation}
\label{Vaidya}
ds^2=\left[1-\frac{2Gm(z)}{c^2 r}\right]dz^2+2dzdr-r^2d\Omega^2,
\label{oldc}
\end{equation}
where $d\Omega^2=d\theta^2+\sin^2\theta d\varphi^2$ is a line element on the 2-dimensional unit sphere. The Vaidya solution describes the metric with a varying mass-function $m(z)$ due to the radial influx at $z=-v$ (or outflux at $z=u$) of null particles \cite{LinSchMis65}. Here $u$ is the advanced time null coordinate and, correspondingly, $v$ is the retarded time null coordinate. This solution is usually expressed either in the $(v,r)$-- or $(u,r)$--coordinates. Later, the Vaidya problem was explored also in the double-null $(v,u)$-coordinates \cite{WauLak86}. 

The Vaidya metric is popular for very different astrophysical and theoretical applications. In particular, it is used for description of evaporating black holes \cite{VolZagFro76,Hiscock81,Kuroda84,Beciu84,Kaminaga90,Zheng94,Farley06,Sawayama06}. The other popular application of Vaidya solution is for the physically reasonable modeling of the relativistic astrophysical objects and cosmological scenarios with radially directed outgoing or ingoing radiation  \cite{Knutsen84,Barreto93,Adams94,Maharaj02,Sungwook10,Alishahiha14}.  This metric is also relevant to the problems of the gravitational collapse, internal black hole structure and formation of naked singularities \cite{HiscockWilliamsEardley,Kuroda84b,Papapetrou85,WauLak86b,PoisIs89,PoisIs89b,PoisIs90,Dwivedi89,Dwivedi91,Joshi92, Joshi92b,Joshi93,Dwivedi95,Ghosh01,Mkenyeleye14}. 

The original Vaidya metric is geodesically incomplete. For better understanding the physical meaning of the Vaydya solution it is important to construct its analytic extension. This extension was done by W. Israel \cite{Israel67} in a rather general form for the global geometry of eternal space-time with the infinite stairs of the Vaidya black-holes and white-holes. Some other approaches for the construction of analytic extension for the Vaidya metric use the combinations of physically reasonable particular subregions \cite{Kuroda84,Fayosyx95} and/or the specific black hole mass functions \cite{WauLak86,HiscockWilliamsEardley,Krori74}. 

The fruitful methods for analytic investigations of the Vaidya metric are the using of the double-null coordinates \cite{WauLak86} and the linear mass function $m(z)$ \cite{VolZagFro76}. In this paper we have found the transformation of Vaidya metric (\ref{Vaidya}) to the new diagonal coordinates, which permit the simple analytical solution for all metric functions in the special case of linear growth of the black hole mass $m(z)$. Our crucial ansatz for finding the maximal analytic extension of the Vaidya metric is in using of the linear black hole mass function, $m(z)=-\alpha z+m_0$, $dm/dz=-\alpha=const$, where parameter $\alpha>0$ is the mass accretion or emission rate. By using this ansatz, we can solve our problem analytically. The Vaidya metric with a linear mass function was extensively investigated earlier by many authors in different aspects (see, in particular, \cite{VolZagFro76,Hiscock81,LevinOri,Shao05,AbdallaChirenti,YangJeng06,BengtssonSenovilla09,ParikhWilczek,Bengtsson}). The novelty of our approach with respect to these previous works is in finding of the special diagonal coordinates, which provide the derivation of the exact analytical expressions for all metric functions. 

The use of specific new coordinates allowed us to reveal the global structure of space-time for Vaidya problem with the linear growth of the black hole mass $m(z)$. Exact solutions for radial photon geodesics provided the main tool for this study. As a result, the global geometry (with physical restriction $m > 0$) was constructed and the corresponding conformal Carter-Penrose diagram was drown. Note that some specific properties of the Vaidya metric in the diagonal coordinates were explored in \cite{LinSchMis65} without derivation of the corresponding metric functions. 

In the Section~\ref{newcoorsec} we explain in details the method of transformation between the used coordinates. Section~\ref{constrsec} is devoted to the detailed construction of the corresponding Carter-Penrose diagrams. Everywhere, except for the Figures, we put $G=c=1$.

%%%%%%%%%%%%%%%%%%%%%%%%%%%%

\section{Transformation to new coordinates}
\label{newcoorsec}

For the detailed investigation of the Vaidya metric with the linear growth of the black hole mass $m(z)$ it would be useful transform the Vaidya metric (\ref{Vaidya}) to the orthogonal frame with some new coordinates, respectively, $\eta$  and $y$:
\begin{equation}
\label{metric-etay}
ds^2=f_0(\eta,y) d\eta^2-\frac{dy^2}{f_1(\eta,y)}-r^2\left(d\theta^2+\sin^2\theta d\phi^2\right),
\label{newc}
\end{equation}
with two principal metric functions $f_0(\eta,y)$ and $f_1(\eta,y)$. 
Note, that the diagonal Vaidya metric was written already in \cite{Vaidya43} (see also Eq.~(9.31) in \cite{GriffithsPodolsky}) with $y=r$ but the particular $g_{\mu\nu}$ calculations require the specification of the form of the mass function $m(z)$, if one goes through the way of transformation from (\ref{oldc}) to diagonal metric (\ref{newc}).

%See the other approaches to the derivation of the diagonal representation of the Vaidya metric in \cite{ParikhWilczek,GriffithsPodolsky,Bengtsson}. (We thanks the anonymous referee of this paper for %indication to these references).

Let us find the transformation between coordinates $(v,r)$ and $(\eta,y)$, i.\,e., from metric (\ref{oldc}) to (\ref{newc}). The first new variable $\eta$ will be defined later, and for the second new variable we choose $y=1-2Gm(z)/r$. By differentiation with respect to $r$ in the relations
\begin{equation}
\label{r}
r=\frac{2m(z)}{1-y}
\end{equation}
and $z(\eta,y)$, we find
\begin{equation}
\label{dr}
dr=\frac{2}{1-y}\frac{dm}{dz}dz+\frac{2m(z)}{(1-y)^2}dy, \quad 
dz=z_{,\eta}d\eta+z_{,y}dy.
\end{equation}
Putting the relations in the two-dimensional part of the metric (\ref{Vaidya}), we derive the system of three equations for functions $f_0$, $f_1$ and $z(\eta,y)$:
\begin{eqnarray}
\label{f0f1}
f_0&=&\left(y+\frac{4}{1-y}\frac{dm}{dz}\right)z_{,\eta}^2, \\
0&=&\left[\left(y+\frac{4}{1-y}\frac{dm}{dz}\right)z_{,y}+\frac{2m}{(1-y)^2}\right]z_{,\eta},\label{eq6} \\
\frac{1}{f_1}&=&-\left[\left(y+\frac{4}{1-y}\frac{dm}{dz}\right)z_{,y}
+\frac{4m}{(1-y)^2}\right]z_{,y}. 
\end{eqnarray}
From the last two equations we find for $dm/dz=-\alpha$:
\begin{equation}
\label{f1}
\frac{1}{f_1}=-\frac{2m}{(1-y)^2}z_{,y}.
\end{equation}
Inserting here the expression for $z_{,y}$ from (\ref{eq6}), we obtain $f_1$ and by substituting it again into (\ref{f1}) we obtain
the differential equation for the function  $z(\eta,y)$:
\begin{equation}
\label{zy}
z_{,y}=\frac{2m}{(1-y)(y^2-y+4\alpha)}.
\end{equation}
Now, multiplying the both parts of this equation by $dm/dz=-\alpha$, we come to the equation, which is possible to integrate directly:
\begin{equation}
\label{dlnm}
\frac{d\log m}{dy}\bigg|_{\eta=const}=-\frac{2\alpha}{(1-y)(y^2-y+4\alpha)}.
\end{equation}
It is very helpful that the solution of equation (\ref{dlnm}) is factorized, $m=C(\eta)\Phi(y)$, with
\begin{equation}
\label{Phi}
\Phi(y)\equiv\exp\left[-2\alpha\int\frac{dy}{(1-y)(y^2-y+4\alpha)}\right]>0.
\end{equation}
As a further step, by differentiating the equation (\ref{r}) with radius $r$, find the expression for derivative
\begin{equation}
\label{y3y4first}
\frac{\partial r}{\partial y}=\frac{2C(\eta)\Phi(y)(y^2-y+2\alpha)}{(1-y)^2(y^2-y+4\alpha)}.
\end{equation}
Here the roots of a square trinomial in the nominator $y^2-y+2\alpha$ equals
\begin{equation}
\label{y1y2}
y_1=\frac{1}{2}(1-\sqrt{1-8\alpha}), \quad y_2=\frac{1}{2}(1+\sqrt{1-8\alpha}).
\end{equation}
At the same time, the roots a square trinomial $y^2-y+4\alpha$ in the integrand in Eq.~(\ref{Phi}) are
\begin{equation}
\label{y3y4}
y_3=\frac{1}{2}(1-\sqrt{1-16\alpha}), \quad y_4=\frac{1}{2}(1+\sqrt{1-16\alpha}).
\end{equation}
If all these roots are real, they are the special points requiring further investigation. Note, that  $0<y_1<y_3\leq1/2\leq y_4<y_2<1$.

By redefinition of time variable  $\eta$ it is always possible to make $C_{,\eta}=const$. Then, for the continuous transition to the Schwarzschild limit $\alpha\to0$ it should be $C(\eta)=\alpha\eta+C_0$ , with $C_0=const$. 

Thus, we get the requested solution in quadratures for the Vaidya metric (\ref{metric-etay}) in the coordinate frame $(\eta,y)$ with a linear black hole mass function $m(z)=-\alpha z+m_0$. Note, that we found the explicit dependence of all metric functions in (\ref{metric-etay}) on coordinates $\eta$ and $y$, not a parametric ones. The integral in (\ref{Phi}) depends on the relation $\alpha\gtreqless1/16$. There are three essentially distinguished cases: (1) the powerful accretion at $\alpha>1/16$, (3) transient case at $\alpha=1/16$ and (3) weak accretion at $\alpha<1/16$. The qualitatively different cases $\alpha>1/16$, $\alpha=1/16$ and $\alpha<1/16$ was identified before in \cite{Bengtsson} and analysed in the non-diagonal coordinates $(v,r)$. We consider below these cases separately in details by using our final formulas:
\begin{equation}
\label{sum}
m=C\Phi, \quad r=\frac{2C}{1-y}\Phi, \quad z_{,\eta}=-\frac{C_{,\eta}}{\alpha}\Phi,
\end{equation}
\begin{equation}
\label{f02}
f_0=-\frac{(y^2-y+4\alpha)}{1-y}\frac{C_{,\eta}^2}{\alpha^2}\Phi^2,
\end{equation}
\begin{equation}
\label{f12}
f_1=-\frac{(1-y)^3(y^2-y+4\alpha)}{(2C\Phi)^2},
\end{equation}
\begin{equation}
\label{Phiagain}
\Phi(y)=\exp\left[-2\alpha\int\frac{dy}{(1-y)(y^2-y+4\alpha)}\right].
\end{equation}

%%%%%%%%%%%%%%%%%%%%%%%%%%%%

\section{Construction of the global geometries}
\label{constrsec}

Our aim is to construct for these 3 cases the corresponding conformal Carter-Penrose diagrams for the corresponding global geometries. 

We define the invariant
\begin{equation}
\label{Y}
Y=\gamma^{ik}y_{,i}y_{,k}=-f_1=\frac{1}{4m^2}(1-y)^3(y^2-y+4\alpha),
\end{equation}
where $\gamma^{ik}$ is the 2-dimensional part of Vaidya metric (\ref{oldc}), without angle variables. We will call the space-time regions with $Y<0$ the ``$R^*$-regions'', and the space-time regions with $Y>0$ the ``$T^*$-regions'', respectively. In the $R^*$-regions $\eta$ --- is the time coordinate and $y$ --- is a space coordinate, within the $T^*$-regions, on the contrary, $\eta$ --- is a space coordinate, à $y$ --- is a time coordinate. 

The physical condition $m>0$ provides the restrictions $y<1$ and $C(\eta)>0$. We remind also that $\Phi>0$ by definition.

It is interesting to note that the described metric is reduced to the simple form by conformal transformation
\begin{equation}
\label{conformalmetric}
ds^2=\frac{C^2}{\alpha^2}\frac{\Phi^2}{(1-y)}\left\{(y^2-2y+4\alpha)
\left[(d\log\Phi)^2)-(d\log C)^2\right]\right\}-r^2d\Omega^2.
\end{equation}
From such a representation it is easily to find for the radial null geodesics:
\begin{equation}
\label{redialnull}
C=A\, \Phi^{\pm1}, \quad A=const.
\end{equation}
In the case of the accretion the upper sign ($+$) corresponds to the outgoing null rays, the lower one ($-$) -- to the ingoing null rays. And vice versa for the emission Vaidya metric.

Taking into account the diagonal form (\ref{conformalmetric}), we draw the Carter-Penrose diagrams in the coordinates $\log C$ and $\log\Phi(y)$ under the following $\arctan$-type transformation 
\begin{eqnarray}
t=\arctan\left[\log C + \log\Phi(y)\right] - \arctan\left[\log C - \log\Phi(y)\right]
\\
x=\arctan\left[\log C  + \log\Phi(y)\right] + \arctan\left[\log C - \log \Phi(y)\right],\nonumber
\end{eqnarray}
with the corresponding shifts and axes exchanges (as needed).

   \subsection{Powerful accretion at $\alpha>1/16$}
   
We start from the simplest case of the powerful accretion when the accretion rate $\alpha>1/16$. Then, $Y>0$, and, therefore, everywhere in the $T^*$-region. In this case the integration in Eq.~(\ref{Phi}) gives us
\begin{equation}
\label{PhiLarge}
\Phi=\frac{\sqrt{1-y}}{(y^2-y+4\alpha)^{1/4}}
\exp\left[-\frac{1}{2\sqrt{16\alpha-1}}\left(\arctan\frac{2y-1}{\sqrt{16\alpha-1}}+\frac{\pi}{2}\right)\right]\!.
\end{equation}

\begin{figure}[h]
		\includegraphics[angle=0,width=0.5\textwidth]{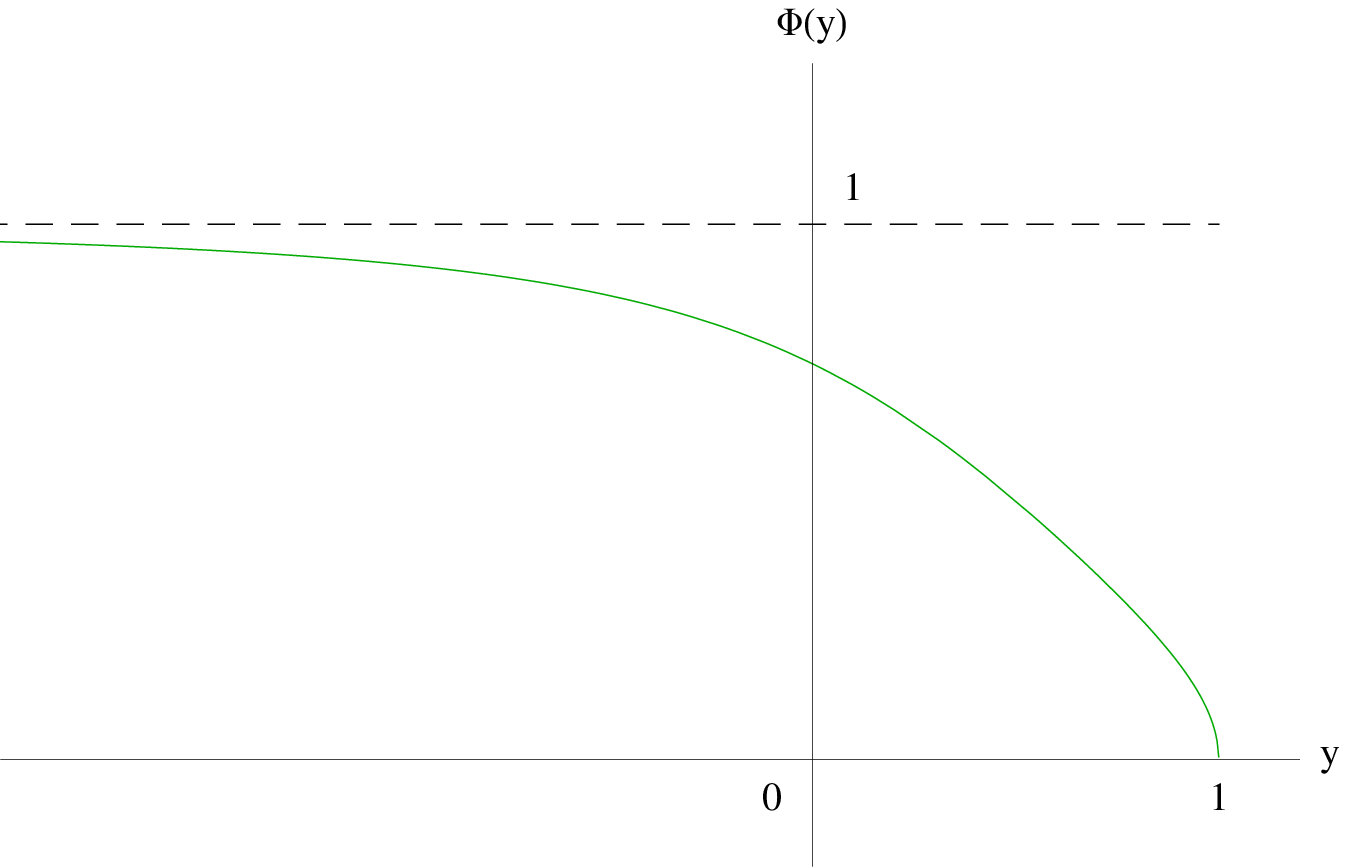}
		\includegraphics[angle=0,width=0.5\textwidth]{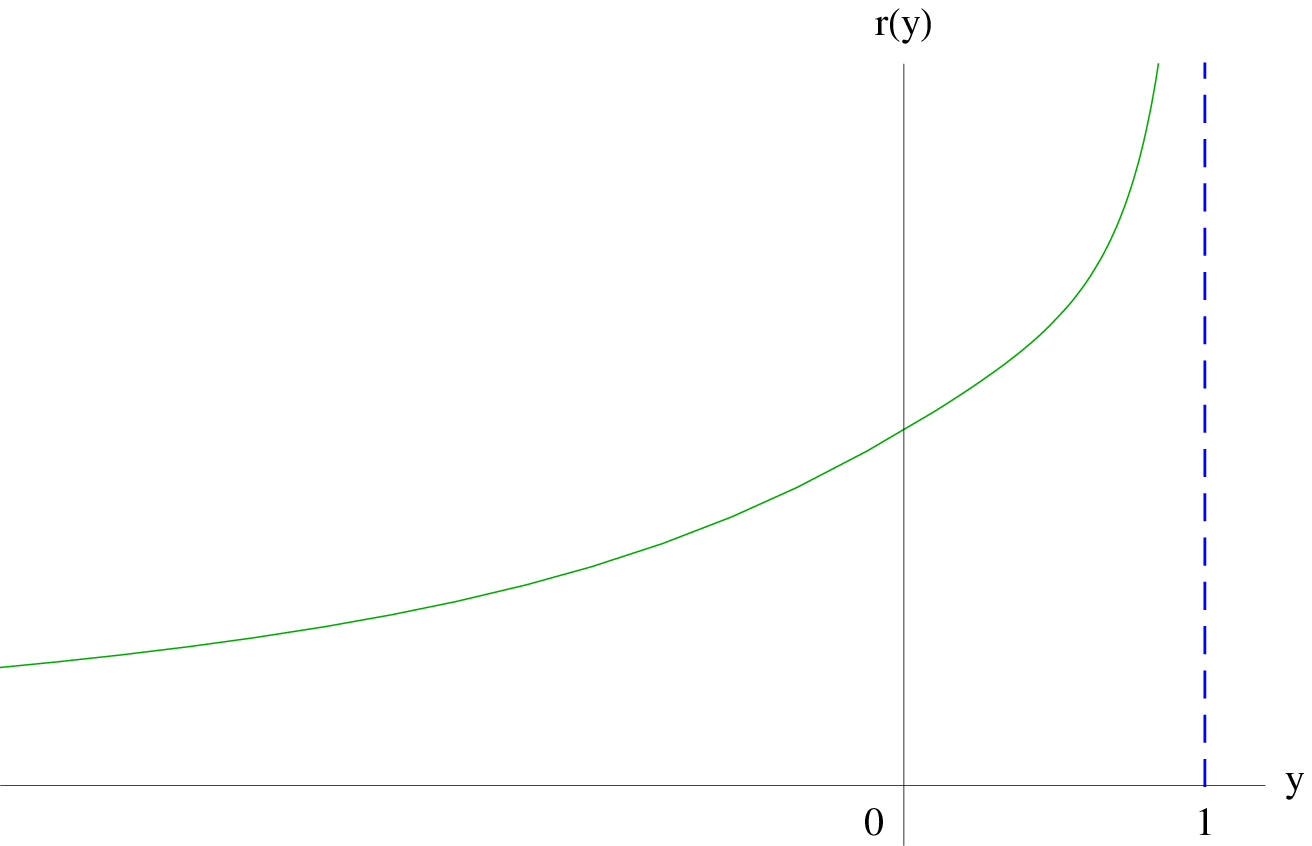}
	\caption{Left: Function $\Phi(y)$ in the case $\alpha>1/16$. Right: Function $r(y)$ in the case  $\alpha>1/8$. }
	\label{Phigtr}
\end{figure}		

\begin{figure}[h]		
		\includegraphics[angle=0,width=0.5\textwidth]{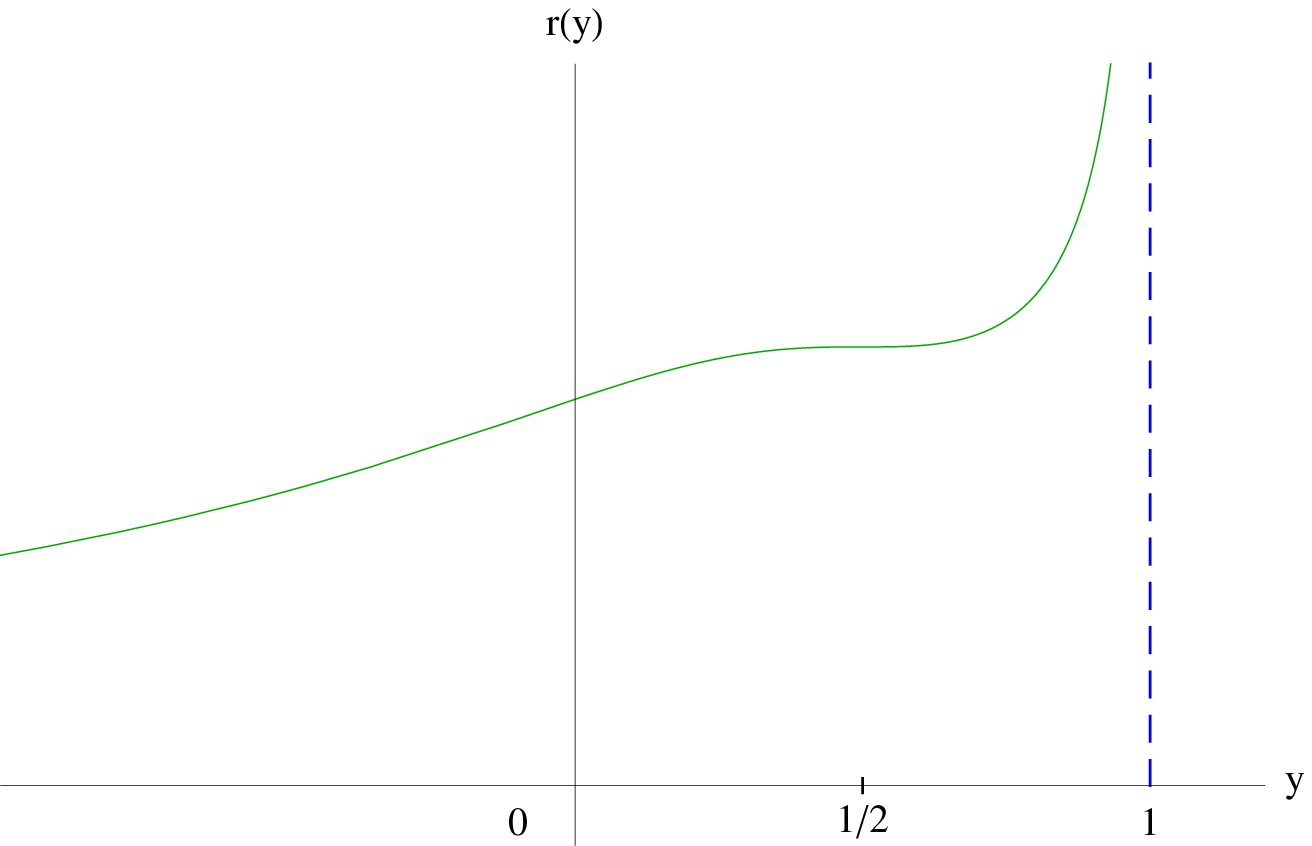}
        \includegraphics[angle=0,width=0.5\textwidth]{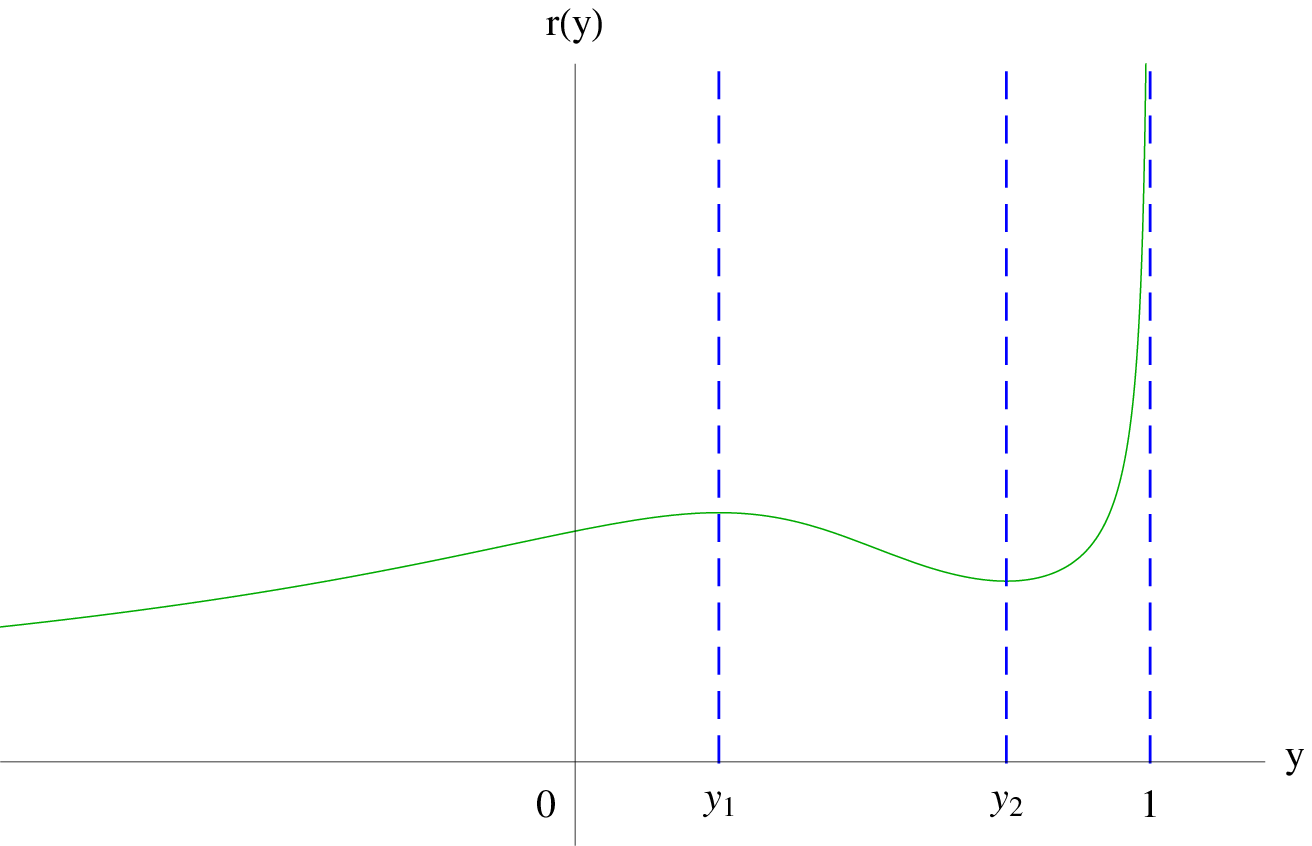}		
	\caption{Left: Function $r(y)$ in the case  $\alpha=1/8$. Right: Function $r(y)$ in the case of a ``just-powerful'' accretion with $1/16<\alpha<1/8$.}
	\label{rgtr18}
\end{figure}

\begin{figure}[h]
	\begin{center}
		\includegraphics[angle=0,width=0.8\textwidth]{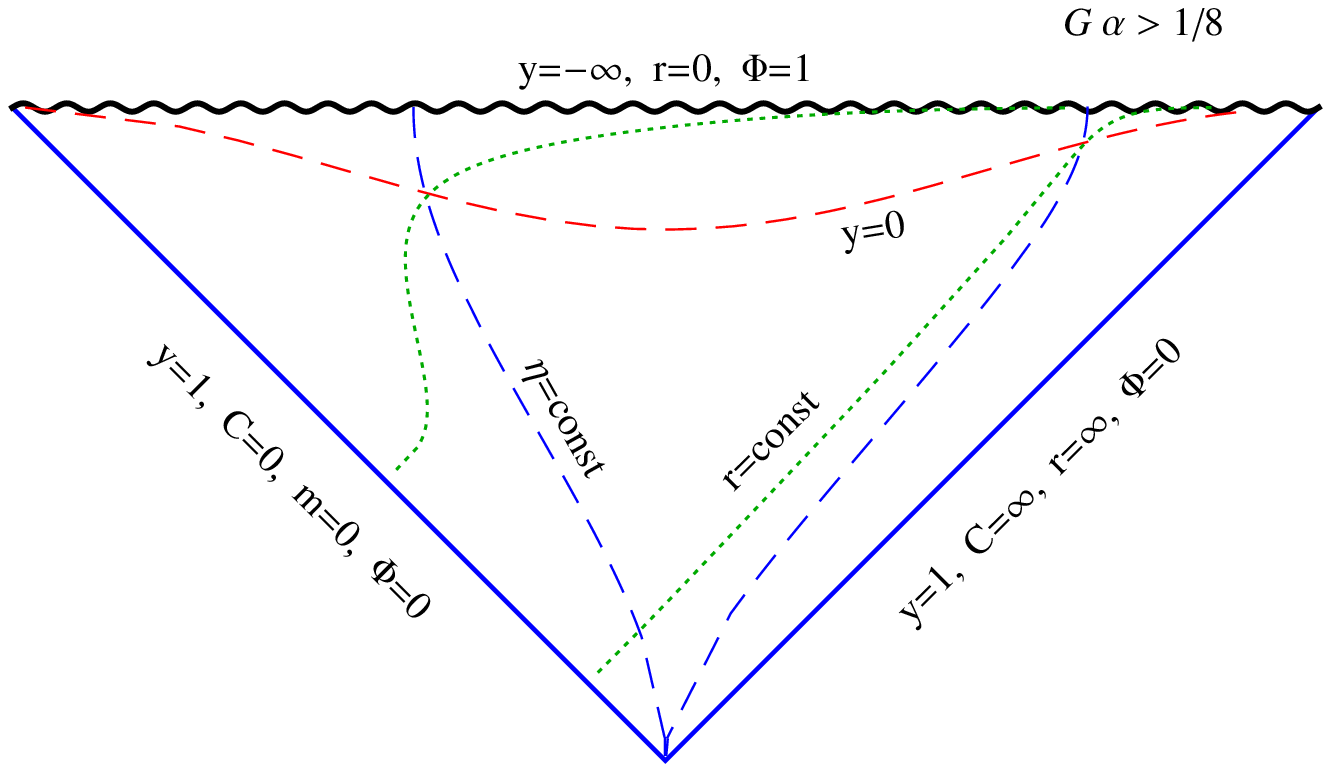}
	\end{center}
	\caption{Carter-Penrose diagram for the global geometry of the Vaidya metric in the case of a ``super-powerful'' linear accretion with $\alpha>1/8$. The boundaries are the horizontal space-line wavy line $y=-\infty$ (in black) corresponding to the curvature singularity at zero radius, $r=0$, and two null lines $y=1$ (in blue), one of them being the past null infinity with $r=\infty$ (it goes in the direction of the advanced time $=$ constant retarded time) while the other corresponds to the marginal initial accretion ray with zero mass, $m=0$ (it goes in the direction of the retarded time $=$ constant advanced time). It is assumed that the time coordinate runs from down to up, and the spatial coordinate --- from left to right. Everywhere inside is the $T^*$-region, i.\,e., the temporal coordinate is ($-y$) running from $y=-\infty$ at the top of the Figure to $y=1$ at the boundaries (the spacelike surfaces $y=const$ are shown by red dashed curves), and the spatial coordinate is $\eta$ (the corresponding timelike surfaces $C(\eta)=const$ are shown by blue dashed curves). The dotted green curves represent the levels of constant radii, $r=const$, they start from the null line $y=1$ (where $m=0$) and end at the right-up corner of the diagram at $y=-\infty$. On the way, they cross the spacelike line $y=0$ that serves as the apparent horizon and separates the space-time region where surfaces $r=const$ are timelike ($y>0$) from that ones where they are spacelike ($y<0$). Evidently, the space-time on the diagram is not geodesically complete because we imposed the physically reasonable condition of the non-negativity of the running black hole mass, $m\geq0$ (remember, that $y=1-2Gm/r$).}
	\label{diagr-superpower}
\end{figure}

\begin{figure}[h]
	\begin{center}
		\includegraphics[angle=0,width=0.8\textwidth]{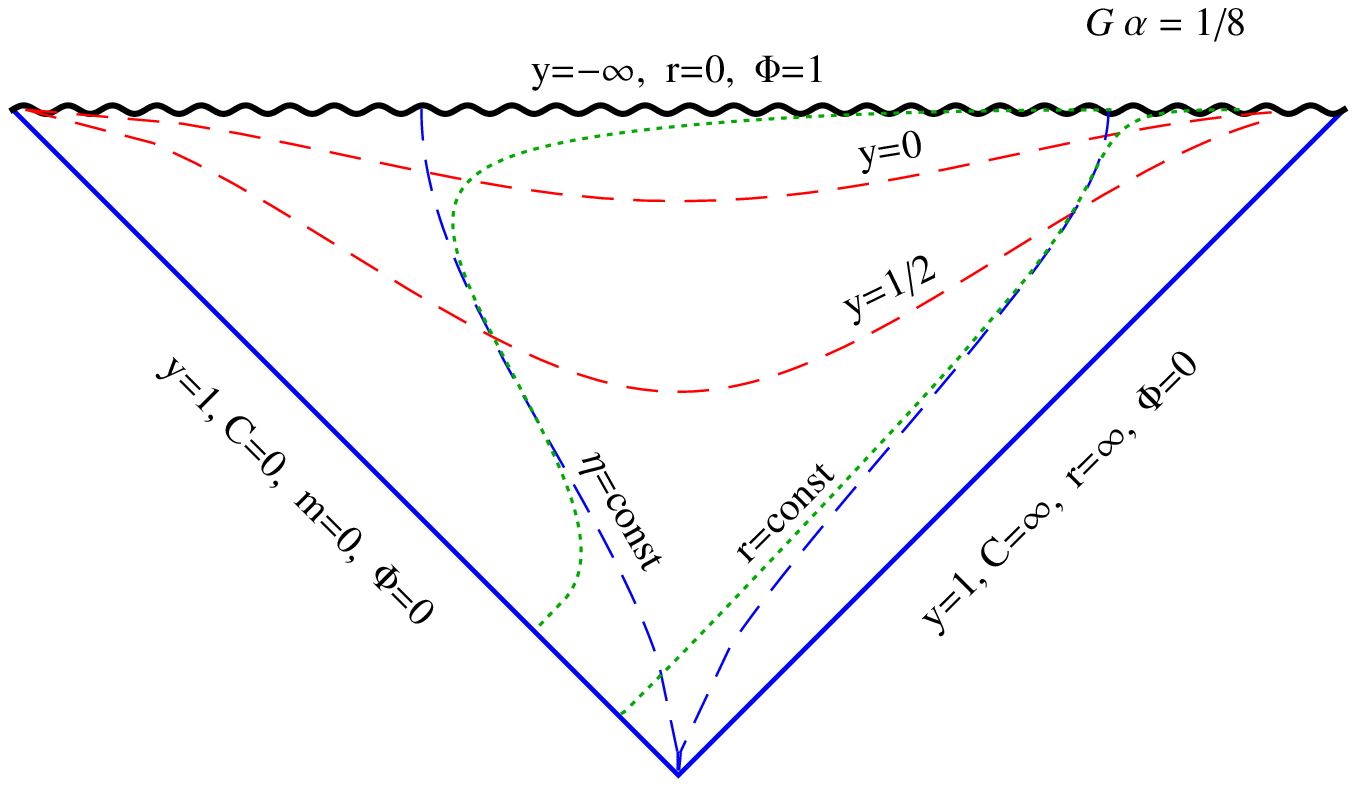}
	\end{center}
	\caption{Carter-Penrose conformal diagram for the transition case from the ``superpowerful'' to ``just-powerful'' accretion, when $G\alpha=1/8$. It differs from the previous one by appearing of the inflection points in the curves $r=const$, corresponding to that ones $y=y_1=y_2=1/2$ in the plots $r=r(y)$.}
	\label{diagr-superpower18}
\end{figure}

\begin{figure}[h]
	\begin{center}
		\includegraphics[angle=0,width=0.8\textwidth]{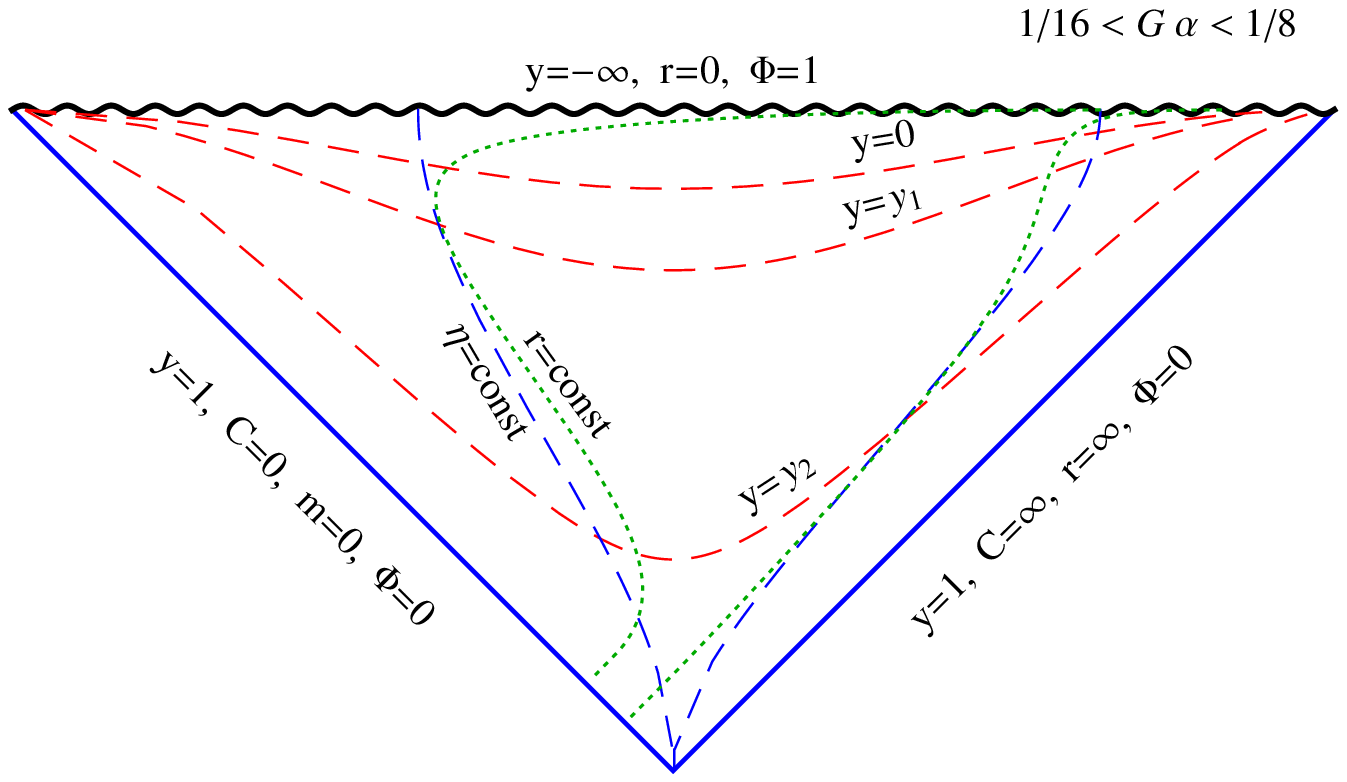}
	\end{center}
	\caption{Carter-Penrose diagram in the case of ``just-powerful'' accretion, $1/16<G\alpha<1/8$. The curves of constant radii, $r=const$, become ``more curvy'' what reflects the appearances of maxima at $y=y_1=(1-\sqrt{1-8G\alpha})/2$ and minima at $y=y_2=(1+\sqrt{1-8G\alpha})/2$ in the plots $r=r(y)$.}
	\label{diagr-power}
\end{figure}

The behavior of the function $\Phi$ is qualitatively the same in the whole range of $\alpha>1/16$. It is shown in Fig.~\ref{Phigtr}. However, the dependence of the radius on $y$, i.e. the function $r(y)$, is different in the intervals $1/16<1/8<\alpha$, $\alpha=1/8$ and  $1/16<\alpha<1/8$. According to the Eqs.~(\ref{y3y4first}) and (\ref{y1y2}), if $\alpha>1/8$, the radius is monotonically increasing function of $y$ (from $r=0$ at $y=\infty$, to $r=\infty$ at $y=1$). We will name this case by the super-power accretion. Then, if $\alpha=1/8$, there appears the point of inflection at $y_1=y_2=1/2$. Finally, if $1/16<\alpha<1/8$, one has the local maximum at $y=y_1=(1-\sqrt{1-8\alpha})/2$ and the local minimum at $y=y_2=(1+\sqrt{1-8\alpha})/2$. We will call this case the ``just-powerfull'' accretion. The corresponding curves are shown in Figs.~\ref{Phigtr},~\ref{rgtr18}. Consequently, the curves of constant radii, $r=const$, will be different on the Carter-Penrose conformal diagrams, shown in Figs.~\ref{diagr-superpower},~\ref{diagr-superpower18},~\ref{diagr-power}. Note also that the curves $y=y_1$, $y=y_2$ (and, of course, $y=y_1=y_2=1/2$) are spacelike.

It is well known that the arbitrary two-dimensional space-time is conformally flat. The Carter-Penrose conformal diagram for the complete two-dimensional flat (Minkowski) space-time is the square with four null infinities as the boundaries. The four-dimensional flat (Minkowskian) space-time written in the spherical coordinates, can be represented, for fixed angles, by the triangle with two null infinities and the timelike (vertical) line $r=0$.  When a spherically symmetric space-time is curved, in its different parts the corresponding conformal factors can be different. Beside the ``natural'' boundaries such as $r=0$ and infinities, there may exist also the various types of horizons (null, timelike or spacelike ones) manifesting the margins of diagram for the spherically symmetric space-time will consists of some set of triangles and squares separated by common boundaries.

The main difference between Fig.~\ref{diagr-superpower} and the Schwarzschild space-time is that the spacelike line $y=0$ represents the apparent horizon and it cannot be an event horizon. Another difference is the construction of infinity. The accretion flow ``drags away'' the coordinate system causing these qualitative deformations of the geometry. 

In the case of the Vaidya metric we impose the additional physical constraint: the total mass should be non-negative, $m\geq0$. Because of this the physical space-time may appear to be geodesically incomplete.

Let us start to construct the boundaries of the Carter-Penrose conformal diagram in the case of powerfull accretion, $\alpha>1/16$. We have 
\begin{equation}
m=C(\eta)\Phi(y), 
\end{equation}
\begin{equation}
r=\frac{2C(\eta)}{1-y}\Phi(y),
\end{equation}
\begin{equation}
Y=\frac{(1-y)^3(y^2-y+4\alpha)}{4C^2(\eta)\Phi^2(y)},
\end{equation}
\begin{equation}
\label{PhiLarge2}
\Phi=\frac{\sqrt{1-y}}{(y^2-y+4\alpha)^{1/4}}
\exp\left[-\frac{1}{2\sqrt{16\alpha-1}}\left(\arctan\frac{2y-1}{\sqrt{16\alpha-1}}+\frac{\pi}{2}\right)\right]\!.
\end{equation}
Note that the expression (\ref{PhiLarge2}) for function $\Phi$ is similar to the function of \cite{Bengtsson} (Eq. (99)-(100)) in the equations for null geodesics, which were studied by \cite{Bengtsson} in the non-diagonal coordinates.

First of all, since, by definition, $\Phi(y)\leq0$, our physical constraint $m\geq0$ leads to the inequalities $C\geq0$, $y\leq1$. It follows from $r\geq0$, that $-\infty<y<1$. Thus, the boundaries are $y=-\infty$ and $y=1$. It is easily seen, that for the powerful accretion, $\alpha>1/16$, everywhere $Y\geq0$, i.e., inside the boundaries we have the so called $T^*$-region, where the curves $y=const$ are spacelike (actually, it is $(-y)$ that plays the role of time, which is supposed to run from down to up), and the curves $\eta=const$ or $C(\eta)=const$ are timelike. Then, we see that the boundary $y=-\infty$ ($r=0$) is spacelike and singular (because the invariant $Y\to+\infty$ at $y\to1$). On the first sight, the boundaries at $y=1$ are null. However, because of $C(\eta)$ and $\Phi(y)$ in the denominator of $Y$, one needs more careful investigation. To do this, we will make use of two congruences of null geodesics, ingoing and outgoing ones. Consider, first, the ingoing rays (the real rays of the accretion) for which $m=C(\eta)\Phi(y)=const$, as it should be. They start from $y=1$, where 
\begin{equation}
r\propto\frac{1}{1-y}, 
\end{equation}
\begin{equation}
Y\propto(1-y)^3 
\end{equation}
and finally encounter the spacelike singularity at $r=0$. We see that, indeed, the boundary is the past null infinity ($y=1$, $r=\infty$). The value of $m$ along this boundary goes from $m=0$ to $m=\infty$, while $\Phi(y)=0$ and $C(\eta)=\infty$.

The nature of the second boundary $y=1$ is a little bit more subtle. Consider the other congruence of null geodesics, for which $C(\eta)=const\cdot \Phi(y)$. They start from $y=1$, where for $y\to1$
\begin{equation}
m\propto(1-y), 
\end{equation}
\begin{equation}
r=const, 
\end{equation}
\begin{equation}
Y\propto(1-y).
\end{equation}
Therefore, this boundary $y=1$ is null, and along it $\Phi(y)=0$, $C(\eta)=0$, $m=0$, and $r$ is running from $r=0$ to $r=\infty$. Therefore, it is no more the infinity, but serves as the margin null ray that starts the accretion. Evidently, such a space-time is not geodesically complete. The corresponding diagrams are presented at Figs.~\ref{diagr-superpower}, \ref{diagr-superpower18}, \ref{diagr-power}.

\begin{figure}[h]
		\includegraphics[angle=0,width=0.5\textwidth]{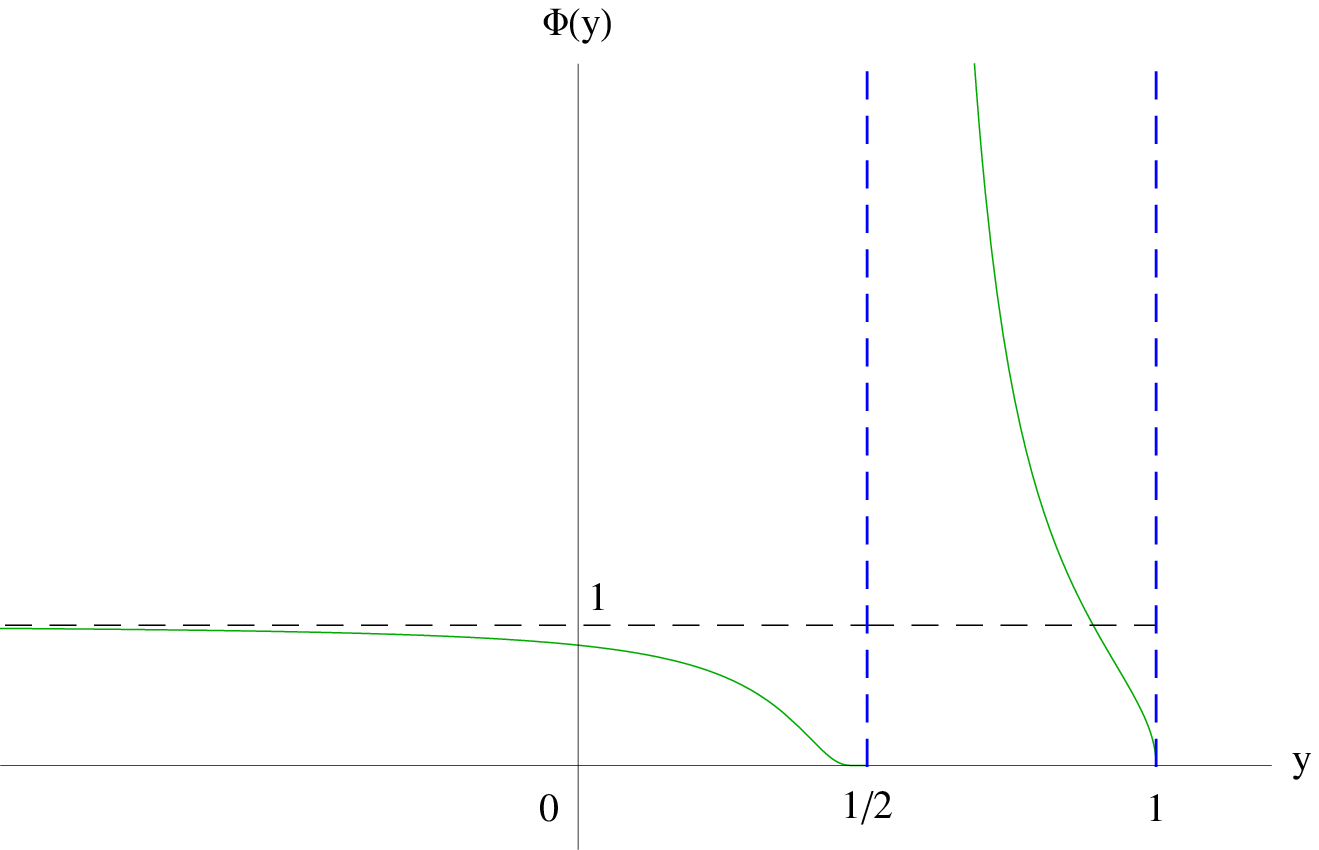}
        \includegraphics[angle=0,width=0.5\textwidth]{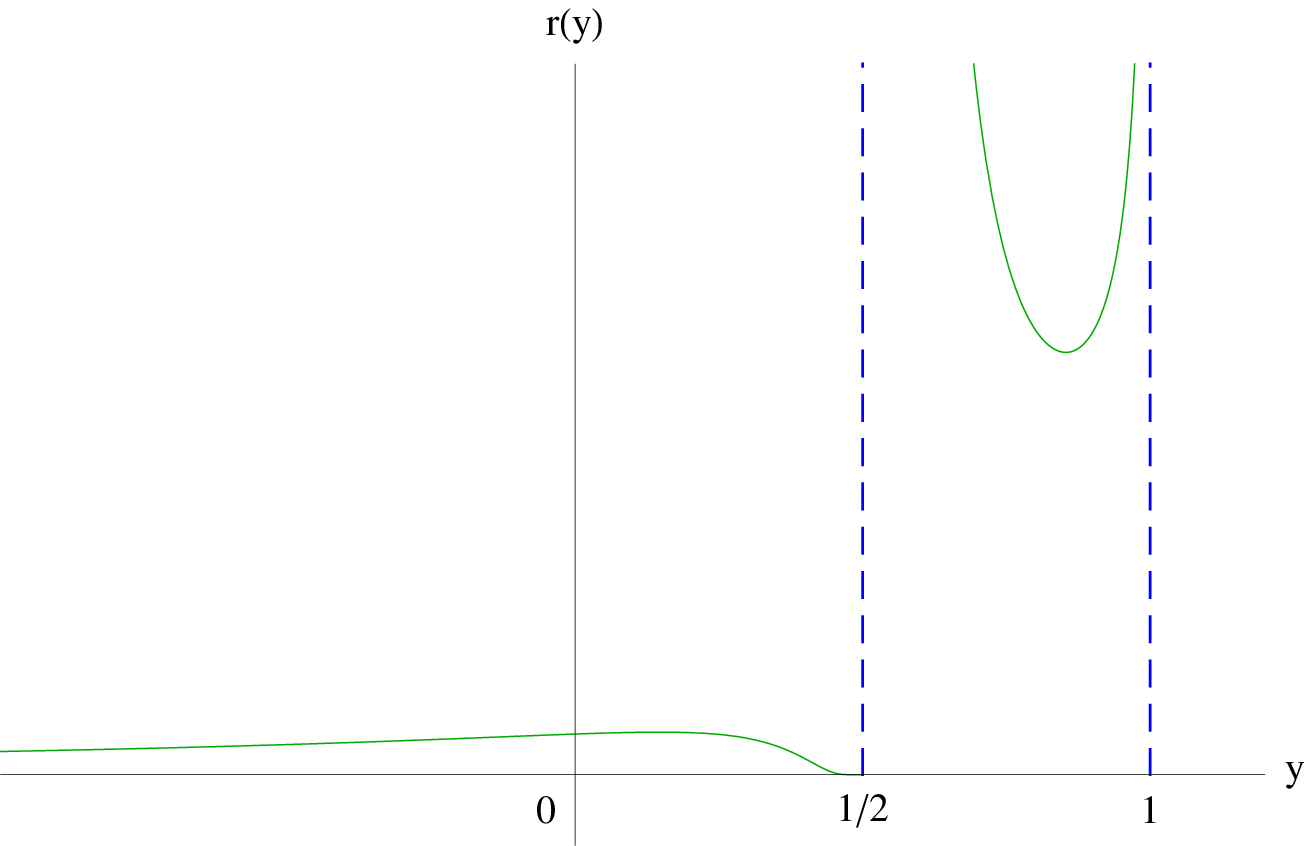}
	\caption{Function $\Phi(y)$ (left) and function $r(y)$ (right) in the transient case between the ``powerful'' and ``weak'' accretion with $\alpha=1/16$.}
	\label{phi-superpower116}
\end{figure}

\begin{figure}[h]
	\begin{center}
		\includegraphics[angle=0,width=0.99\textwidth]{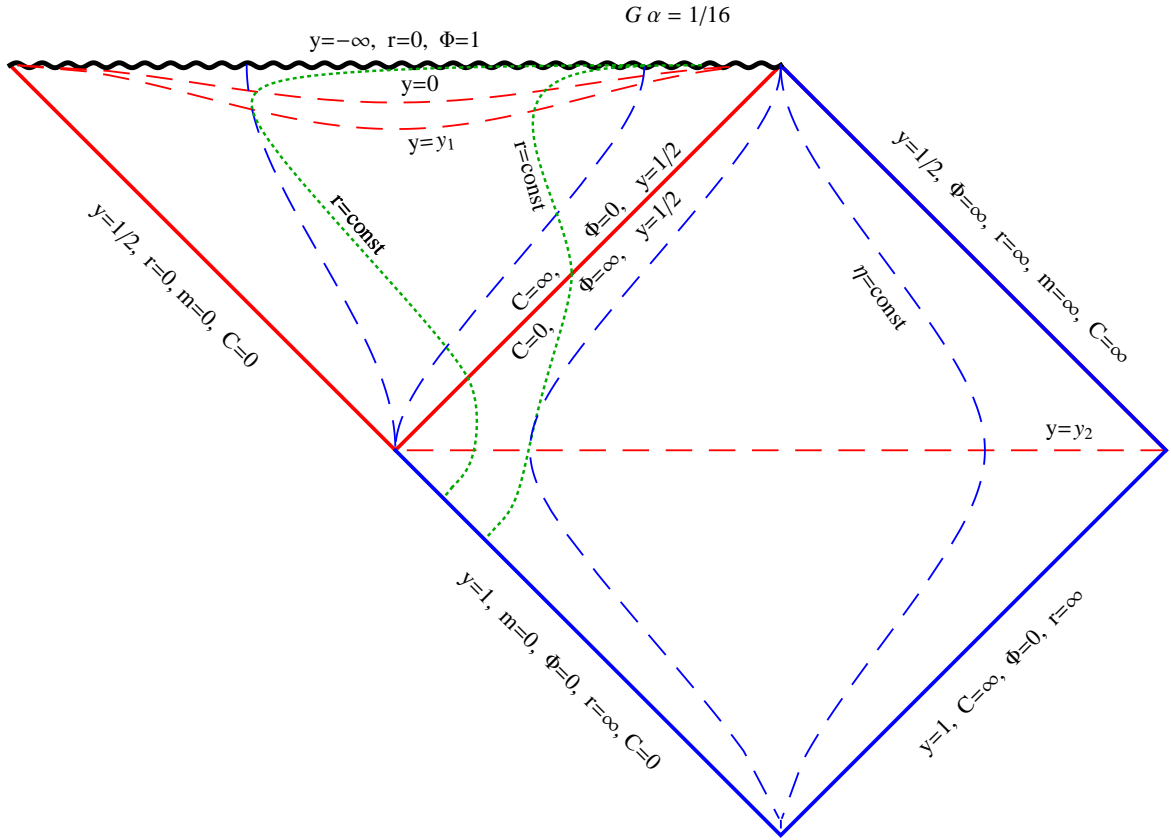}
	\end{center}
	\caption{Carter-Penrose conformal diagram for the transient case between ``powerful'' and ``weak'' accretion, $\alpha=1/16$. New feature is the appearance of the double horizon at $y_3=y_4=1/2$, separating the two $T^*$-regions. The null boundaries are drastically changed. One of them, representing the marginal initial ray, consists now of two parts, the first one where $y=1/2$ with both mass and radius are zero, $m$,$r=0$, and the second, where $y=1$, $m=0$, and running radii $0<r<\infty$. The past null infinity, where $y=1$, $r=\infty$ remains the same. But there appears the future null infinity, where $y=1/2$ with both mass and radius are infinite, $m$,$r=\infty$.}
	\label{diagr-superpower116}
\end{figure}

\begin{figure}[h]\includegraphics[angle=0,width=0.5\textwidth]{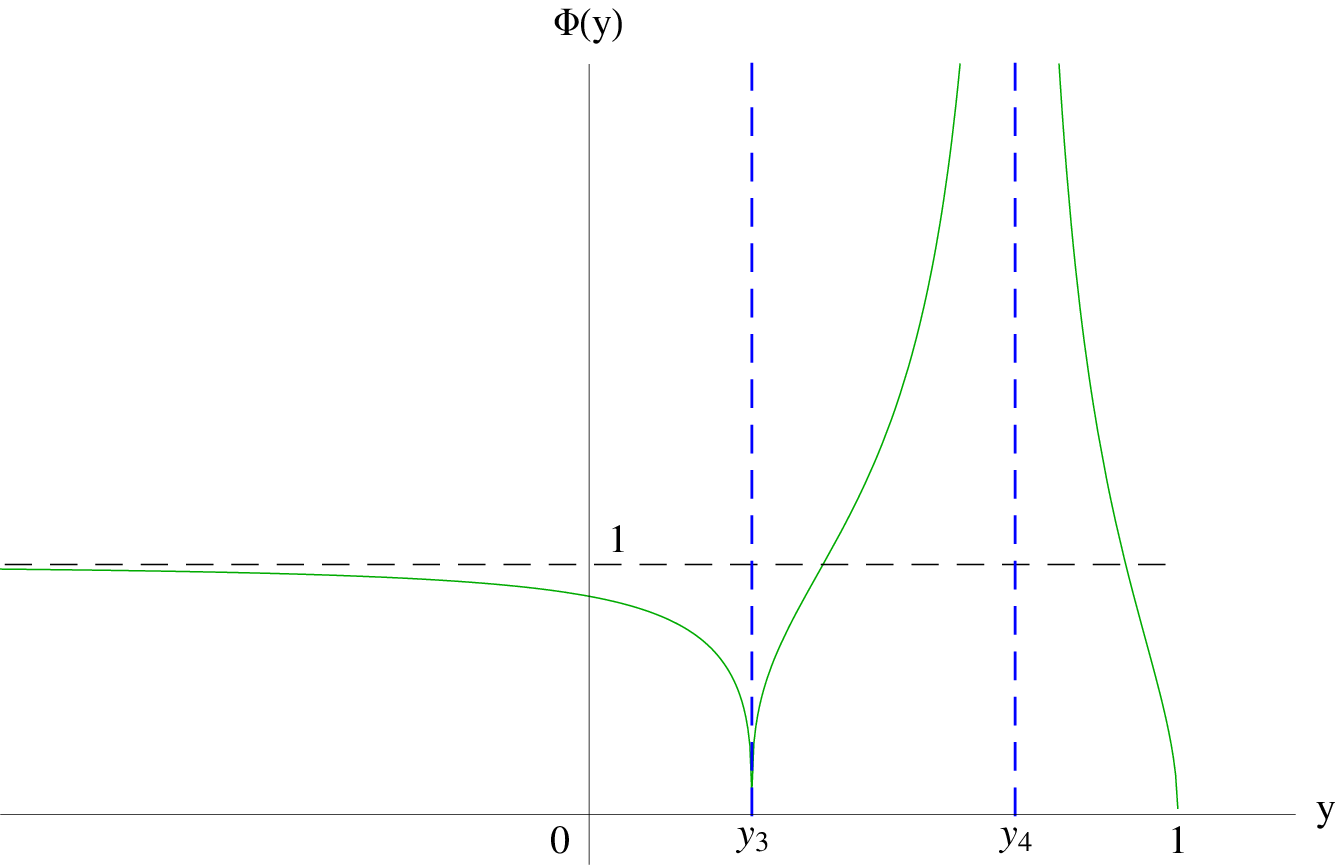}
\includegraphics[angle=0,width=0.5\textwidth]{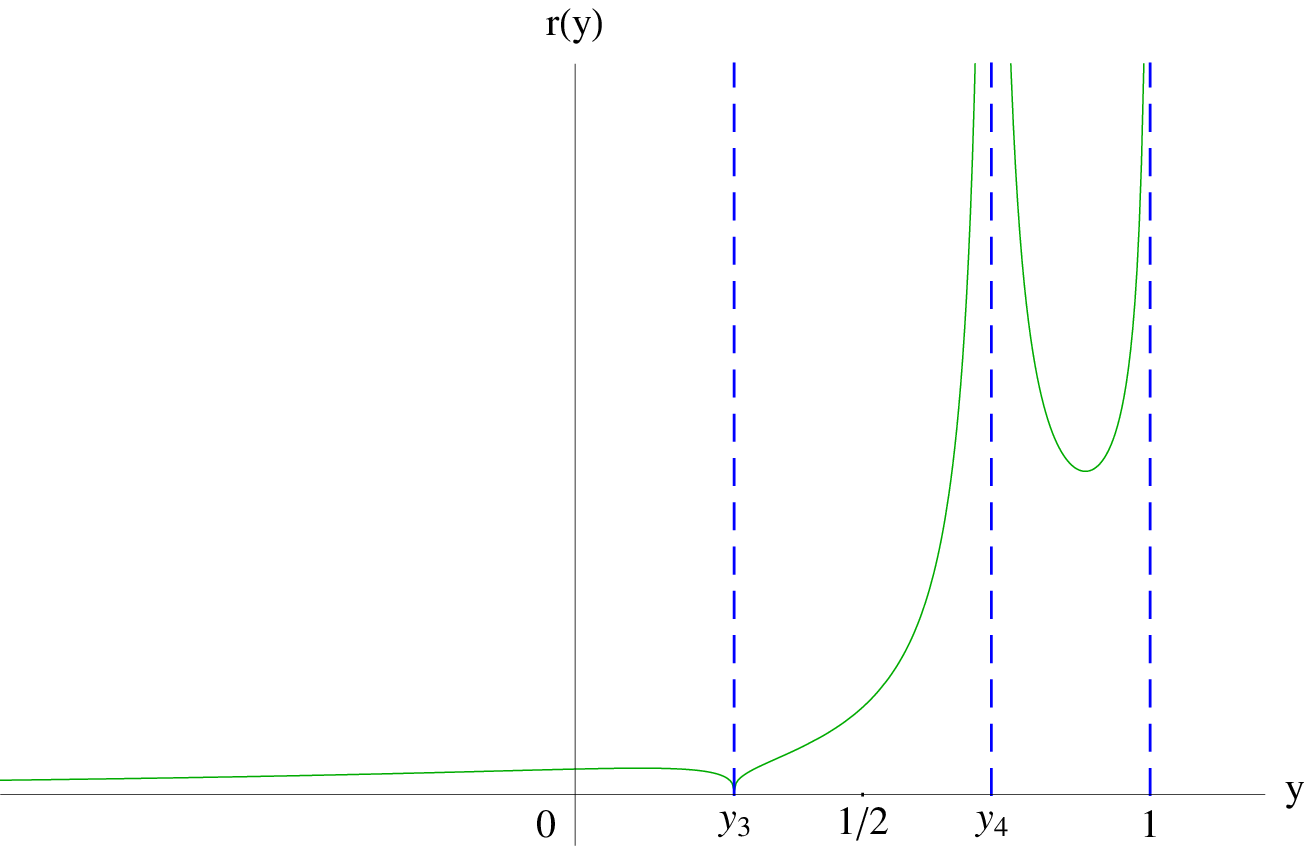}
	\caption{Function $\Phi(y)$ (left) and function $r(y)$ (right) in the case of a ``weak'' accretion with $\alpha<1/16$. }
	\label{phi-slow}
\end{figure}

\begin{figure}[h]
	\begin{center}
		\includegraphics[angle=0,width=0.99\textwidth]{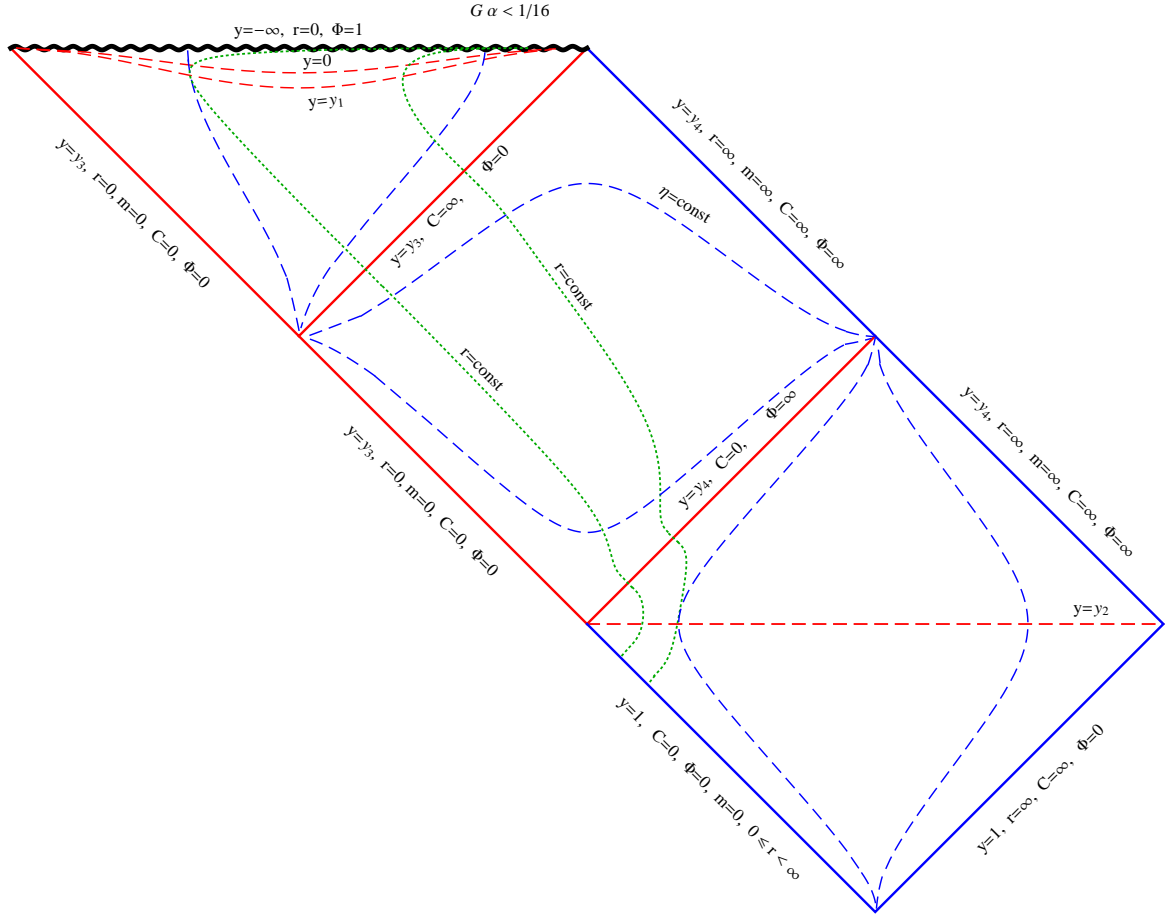}
	\end{center}
	\caption{Carter-Penrose diagram for Vaidya metric in the case of ``weak'' accretion, $G\alpha<1/16$. The global geometry is now the most complex. The double horizon is splitted into two, the event horizon at $y=y_3=(1-\sqrt{1-16G\alpha})/2$ and the horizon at $y=y_4=(1+\sqrt{1-16G\alpha})/2$ (note that $y_1<y_3<y_4<y_2$). One has the $T^*$-regions for $-\infty<y<y_3$ and for $y_4<y<1$. In between, $y_3<y<y_4$, there lies the $R^*$-region where $\eta$ is the time coordinate, where $y$ is the spatial coordinate. The first of the $T^*$-regions is bounded by the spacelike curvature singularity $y=-\infty$, $r=0$, the marginal null ray where $y=y_3$, and both mass and radius are zero, $m,r=0$, and by the event horizon $y=y_3$. The $R^*$-region is bounded by two horizons, $y=y_3$ and $y=y_4$, by the marginal null ray $y=y_3$, $m=0$, $r=0$, and by the future null infinity, where $y=y_4$, $m=\infty$, $r=\infty$. The second $T^*$-region is bounded by the infinity $y=y_4$, the marginal null ray $y=y_1$ with zero mass, $m=0$ and running radii, $0<r<\infty$, by the past null infinity with $y=1$, $r=\infty$ and running mass, $0<m<\infty$, and, finally, by the future null infinity where $y=y_4$ and both mass and radius are infinite, $m,r=\infty$.}
	\label{diagr-slow}
\end{figure}

%%%%%%%

   \subsection{Transient case at $\alpha=1/16$}
   
Now, we have for the function   $\Phi(y)$:
\begin{equation}
\label{Phi2}
\Phi=\sqrt{\left|\frac{y-1}{y-(1/2)}\right|}\exp\left\{\frac{1}{4[y-(1/2)]}\right\},
\end{equation}
and 
\begin{equation}
m=C(\eta)\Phi(y), 
\end{equation}
\begin{equation}
r=\frac{2C(\eta)}{1-y}\Phi(y),
\end{equation}
\begin{equation}
f_0=\left|y-\frac{1}{2}\right|\exp\left\{\frac{1}{2\left(y-\frac{1}{2}\right)}\right\},
\end{equation}
\begin{equation}
Y=-f_1=\frac{(1-y)^3(y-\frac{1}{2})^2}{4C^2\Phi^2}=\frac{(1-y)^2\left|y-\frac{1}{2}\right|^3}{4C^2}\exp\left\{-\frac{1}{2\left(y-\frac{1}{2}\right)}\right\}.
\end{equation}
The behavior of $\Phi(y)$ and $r(y)$ is shown in Fig.~\ref{phi-superpower116}.
The Carter-Penrose diagram consists now of two parts: the triangle and the square, glued together (and separated) by the double horizon $y=y_3=y_4=1/2$. On both sides we have the $T^*$-regions (the curves $y=const$ are spacelike).

The triangle is constructed by the spacelike (horizontal) singular line $y=-\infty$ (as before, on the upside), where $r=0$, and (on the down side) by two null boundaries $y=y_3=y_4=1/2$. One of them is the above mentioned double horizon, while the other is the margin null ray with $m=0$ and $r=0$ that starts the accretion. There is a subtle when dealing with the  double horizon. The matter is that $\Phi(1/2-0)=0$, and $\Phi(1/2-0)=\infty$. This problem can be resolved by considering (again) the accretion rays with $C\Phi=m=const$. We see that $C(\eta)=\infty$ at $y(1/2-0)$, and this is just the end of the spatial ($\eta$!) coordinate range in the triangle. At $y(1/2+0)$ we have $C(\eta)=0$, and this is just the beginning of new spatial  ($\eta$!)  coordinate range in the square. For the marginal ray with $m=0$ along $y=1/2$ both $C$ and $\Phi$ are zero. The divergence of the invariant $Y$ should not confuse us since the point $y=1/2$ is the coordinate singularity (the metric determinant is zero there). Evidently, such a space-time is not geodesically complete. And now, look at Fig.~\ref{diagr-superpower116} for the corresponding Carter-Penrose conformal diagram for the transient case between ``powerful'' and ``weak'' accretion, $\alpha=1/16$.

%%%%%%%

   \subsection{Weak accretion at $\alpha<1/16$}
   
In this case the double horizon is splitted in two at  $y=y_3=(1-\sqrt{1-16\alpha})/2$ and $y=y_4=(1+\sqrt{1-16\alpha})/2$. The very appearance of these ``new horizons'' is due to the fact that we consider here the unbounded accretion which infinitely growing the black hole mass. The function $\Phi(y)$ takes now the form
\begin{equation}
\label{Phiweak}
\Phi=\sqrt{1-y}\,|y-y_3|^{y_3/[2(y_4-y_3)]}\,|y-y_4|^{-y_4/[2(y_4-y_3)]},
\end{equation}
and
\begin{equation}
m=C(\eta)\Phi(y), 
\end{equation}
\begin{equation}
r=\frac{2C(\eta)}{1-y}\Phi(y),
\end{equation}
\begin{equation}
f_0=-|y-y_3|^{\frac{y_3}{y_4-y_3}+1}|y-y_4|^{\frac{-y_4}{y_4-y_3}+1},
\end{equation}
\begin{equation}
Y=-f_1=\frac{(1-y)^2(y-y_3)(y-y_4)}{4C^2}|y-y_3|^{\frac{-y_3}{y_4-y_3}}|y-y_4|^{\frac{y_4}{y_4-y_3}}.
\end{equation}
The curves $\Phi(y)$ and $r(y)$ are shown at Fig.~\ref{phi-slow}.   

The global geometry is more sophisticated. The Carter-Penrose conformal diagram consists of one triangle and two squares. The structure of the triangle boundaries remains exactly the same (except that now $y<1/2$), and we have there the $T^*$-region (with $\eta=const$ timelike and $y=const$ spacelike). But there appears also the $R^*$-region (absent before), with $\eta=const$ spacelike and $y=const$ timelike inside the square, boundaries of which are the ``new horizons'' $y=y_3$ and $y=y_4$. The left hand side boundary consists of two branches, upper and lower, both at $y=y_3$. The latter is just the ``first'' null accretion ray with $m=0$, $C(\eta)=0$, $\Phi=0$ and $r=0$, while the upper one, with $C=\infty$, $\Phi=0$, is the event horizon matching (and separating) the $R^*$-region and the inner $T^*$-region ($y<y_3$). The right hand side boundary also consists of two branches, both at $y=y_4$. The upper one is just the ``last'' accretion ray with $r=\infty$, $m=\infty$, $C(\eta)=\infty$ and $\Phi=\infty$ -- this is the future null infinity, while the lower one is the cosmological horizon matching (and separating) the $R^*$-region and the outer  $T^*$-region ($y_3\leq y_4\leq1$). The second square ($y_4\leq y\leq1$) has the same structure as before, except that now $y_4>1/2$. 

There are two null lines $y=y_4$ on our Carter-Penroze diagram. That one separating the $R^*$- and $T^*$-regions is similar to the cosmological horizon in the Schwarzschild-de Sitter space-time, while the other one corresponds to the future null infinity ($r\to \infty$).

Evidently, such a space-time is not geodesically complete. See Fig.~\ref{diagr-slow} for the corresponding Carter-Penrose diagram for Vaidya metric in the case of ``weak'' accretion, $G\alpha<1/16$.

%%%%%%%%%%%

\section{Conclusion}

For exploring of the global geometry for Vaidya problem with a linear growth of the mass function we found the appropriate coordinate transformation from usual Vaidya metric to the diagonal one. The advantage of this linear model is the possibility for finding of all metric functions and photons geodesics in the exact analytical forms. This coordinate transformation is useful for construction of the Carter-Penrose conform diagrams for the corresponding maximum analytic extension of the Vaidya metric with a linear growth of the black hole mass in the different cases, depending on the value of the energy accretion rate. The corresponding level lines for the temporal and space coordinates, photons geodesics were defined for these diagrams as well as the lines $r=const$. The constructed Carter-Penrose diagrams contain quite different space-time regions, separated by horizons and marginal lines of coordinate systems. The space-time on the constructed diagrams geodesically incomplete because we imposed the physically reasonable condition of the non-negativity of the running black hole mass, $m\geq0$.

\ack

This work was partially supported by the Russian Foundation for Basic Researches (RFBR), research project $15-02-05038$. We thanks also the anonymous referee of this paper for indication to us the references \cite{ParikhWilczek,Bengtsson,GriffithsPodolsky}.

\end{document}